\documentstyle[prb,aps,twocolumn,epsf]{revtex}

\begin{document}
\draft
\preprint{\today}
\title{Prediction of Ferromagnetic Ground State of NaCl-type FeN}
\author{A. Filippetti and W. E. Pickett}
\address{Department of Physics, University of California -- Davis,
Davis, Ca 95616} 
\maketitle
%
\begin{abstract}
Ab initio results for structural and electronic properties of NaCl-type
FeN are presented. Calculations are performed within a plane-wave and
ultrasoft pseudopotential framework.
Competition among paramagnetic, ferromagnetic and some
possible antiferromagnetic phases is examined. We find the ferromagnetic
phase stable overall as observed. Stabilization over the unpolarized 
phase is obtained by splitting one flat $t_{2g}$-type band crossing 
the Fermi energy. Comparison with CrN reveals 
that the magnetic ordering of CrN, which  consists of double-ferromagnetic 
sheets compensating along the [110] direction of the cubic cell and 
stabilized by the distortion of the squared symmetry on (100) plane, 
is not effective for FeN. The different behavior of FeN and CrN can be 
traced back to the smaller ionicity and magnetization of the former.

\end{abstract}

\narrowtext 

\section{Introduction}

Stable compounds\cite{hans} formed by Fe and N, such as 
$\gamma^{\prime}$-Fe$_4$N
and $\zeta$-Fe$_2$N, or quasi-stable\cite{kt} like 
$\alpha^{\prime\prime}$-Fe$_{16}$N have been known for some time. 
On the other hand, almost nothing was reported about FeN until the work 
of Heiman {\it et al.}\cite{hk}, where for the
first time, they synthesized Fe$_x$N$_{1-x}$ with 0.4 $<x<$ 0.75.
In this work FeN was found to be ferromagnetic. Successive studies found
FeN in NaCl-type crystal structure\cite{obff}, and non-ferromagnetic at
room temperature\cite{syf}. More recent works are in contradiction with 
the earlier ones: Morita {\it et al.}\cite{myk} reported FeN antiferromagnet 
at temperatures below 100 K. Moreover, Suzuki {\it et al.}\cite{smkyf} 
determined the FeN structure to be of ZnS fcc type. They also found their
sample to exhibit a mictomagnetic character that was suggested be 
related to the antiferromagnetism of FeN.

Regarding the structure, it seems clear that FeN can assume both zincblende
and rocksalt structure, depending on the sample history, and both of
them may contain some deficiencies of N atoms. Regarding magnetic properties,
recent M\"ossbauer spectroscopy measurements suggested that NaCl-type FeN 
is antiferromagnetic\cite{nntk,hn}, with N\'eel temperature around room 
temperature. However, M\"ossbauer spectroscopy is not able to determine 
completely magnetic ordering. 

For both structural and magnetic properties
ab initio calculations are suited to furnish a prediction about the
ground state of FeN, and to help guide understanding of experimental
results. Here we investigate the relative stability of paramagnetic (PM),
ferromagnetic (FM) and some possible antiferromagnetic (AFM) phases of
NaCl-type FeN that have been observed or suggested for other transition-metal 
nitride (or oxide or carbide) compounds. Compared to other 
transition-metal nitrides (TMNs), it has been noticed\cite{sss} 
that NaCl-type FeN
shows electronic properties close to those of CrN that we recently 
investigated\cite{fpk}, e.g. the Fermi energy falls exactly on a pronounced
peak of the density of states, mainly formed by Fe $d$-like flat bands.
In Ref. \onlinecite{sss} the authors come to the conclusion, on the basis
of the rigid band model to evaluate the DOS and the Stoner condition
of ferromagnetism, that only CrN, FeN and CoN in rocksalt phase are likely 
to be stabilized by magnetic ordering. 

Comparison with CrN is particularly
interesting because it undergoes a cubic-to-orthorhombic transition 
accompanying the PM to AFM ordering at a N\'eel temperature close to
the room temperature, with AFM phase formed by double ferromagnetic sheets
compensating along the [110] direction\cite{}. The structural distortion 
is decisive in determining the most stable among the different AFM phases, 
and stress relief is likely to be the ultimate driving force towards 
distortion. If such an involved mechanism of magnetic-ordering is driven
by the high DOS at Fermi level, it could be argued that a similar
magnetic ordering occurs for FeN as well. Thus, for possible AFM
phases, we decide to consider here the [110]-double and single sheet AFM 
arrangements (indicated as AFM$^2_{[110]}$ and AFM$^1_{[110]}$, respectively
and described fully in Ref.\cite{fpk}),
as well as the AFM$_{[111]}$ phase (very common for NaCl-type transition metal 
oxides), consisting of single ferromagnetic sheets alternating along the 
[111] direction.

Equilibrium lattice constants,
energies, and magnetic moments are presented for all the considered phases.
Also, band structures and density of states (DOS's) are presented for
PM and FM phases.
Our local-spin-density calculations are performed in a plane-wave and 
ultrasoft pseudopotential framework\cite{van}. Use of ultrasoft 
pseudopotentials allow us to obtain well converged results for a cutoff
energy equal to 30 Ryd. We used sets of 10 to 110 special k-points 
(depending on the structure) for the self-consistent calculations, 
up to 280 k-points to evaluate the DOS.

\section{Structure}
Experimentally it was found for the lattice constant of FeN
in rocksalt and zincblende phase $a^{rs}_0$=8.5 bohr\cite{nntk} 
and $a^{zb}_0$=8.1-8.2 bohr \cite{nntk,smkyf}, respectively. This result is
somewhat surprising; indeed, assuming the same anion-cation bond length
in the two phases, the relation between lattice constants would be 
$a^{rs}_0=(\sqrt{3}/2)a^{zb}_0 < a_0^{zb}$. Although this crude 
approximation considerably overestimates the actual difference between 
the two structures, it is 
generally true that in most cases the rocksalt phase have larger volume.
Doubts about the measurements are confirmed by a recent theoretical 
work\cite{sss}
where structural properties for transition metal nitrides are
calculated. They found for $a^{zb}_0$ a value in good agreement with 
experiments, but a much smaller $a^{rs}_0\sim$ 7.5 bohr in the paramagnetic
phase. The discrepancy with experiment is so large that it seems unlikely
it could be due to different magnetic orderings. Also, in 
Ref.\onlinecite{sss} the equilibrium lattice constant for other phases
(ferromagnetic and a couple of antiferromagnetic phases)
was evaluated, but no relevant changes on $a^{rs}_0$ were found.
Finally, the value $a^{rs}_0\sim$ 7.5 bohr is consistent with that of
other rocksalt TMN structures. For instance, experiments assign to AFM
CrN a value $a^{rs}_0\sim$ 7.8 bohr, and we should expect for FeN a 
somewhat smaller lattice constant. 

To try to explain the disagreement with experiments we consider
here other AFM phases. In particular, among the AFM phases examined in 
Ref.\onlinecite{sss}, the AFM$_{[110]}$ was not taken into account. 
The PM to AFM$^2_{[110]}$ transition is found to produce, for CrN, a large 
increase of lattice constant\cite{fpk} ($\sim$ 2\%), although not large enough 
to account for the discrepancy with experiments of FeN,
and the largest energy gain with respect to the PM phase. 
In Table \ref{struct} we report our calculated values of $a^{rs}_0$ for 
the corresponding magnetic orderings. Evidently, contrary to what happens 
for CrN, the structure of FeN is basically unaffected 
by magnetic ordering, and only a 0.4\% increase of $a^{rs}_0$ is found
in the FM phase with respect to the PM. Our results are in substantial 
agreement with Ref.\onlinecite{sss} (they found $a_0$=7.47 and 7.54 for
PM and AFM phases, respectively). 

From our calculations, it results that NaCl-type FeN is ferromagnetic.
The energy gain of FM phase with respect to the PM is less than 0.1 eV 
per formula unit, i.e. significantly smaller than in case of CrN. 
Also, the competition between FM and the
examined AFM phases is very close (energy differences are few hundredths 
of eV). To be confident about the precision of our calculations, we 
evaluated the FM-AFM energy differences by performing FM calculations 
in the same symmetry of all the AFM structures considered, and tested 
the differences for sets of special K-points of increasing size. 
We conclude that, among the AFM phases, the lowest in energy is 
AFM$^1_{[110]}$, also the lowest among the cubic phases of CrN\cite{fpk}. 
In that case, the orthorhombic [110] shear distortion of CrN 
produces a further energy gain for the AFM$^2_{[110]}$ structure 
(anisotropic in the (100) plane due to magnetic ordering) that causes 
the latter to be the most stable phase overall. To investigate the effect of 
planar distortion on FeN we applied the same distortion experimentally 
observed for CrN (i.e. a $\sim$2\% reduction of the bisection angle at the 
squared base of the tetragonal cell. Surprisingly, we find very little
change in the energy with respect to the cubic AFM$^2_{[110]}$ phase (the 
increase is just $\sim$ 2 meV). 

Thus, notwithstanding the similarity of DOS at Fermi level, CrN and FeN 
are stabilized by very different magnetic order and structure. In Table 
\ref{struct} we report the magnetization $m$ (per formula unit, i.e. per 
couple Fe-N) for the considered phases of FeN. The FM phase has the highest 
magnetization (1.65 $\mu_B$), of which $\sim$ 0.15 $\mu_B$ comes from 
polarization of N, and $\sim$ 1.5 $\mu_B$ from polarization of Fe. 
For CrN the magnetization is 
more that 2 $\mu_B$ per formula unit. Also, the charge transfer is
different. For FeN it is equal to about one electron, 
so that the approximate ionic configuration Fe$^{+1}$N$^{-1}$ can be 
deduced, i.e. Fe is near a $d^7$ configuration. In case of CrN, the charge 
transfer is larger (close to 2 electrons), due to the higher electronegativity
of Fe with respect to Cr. i.e. to the smaller electronegativity difference 
with N. The higher covalency of FeN contributes to it having a smaller 
lattice constant, whereas CrN has larger ionicity and polarization. 

In Ref.\onlinecite{fpk} it is argued that for CrN the energy gain from
the planar distortion is due to a relief of tensile stress stored in the
bonds between metal atoms with antiparallel spins. This stress can be 
conveniently defined as the excess stress produced by a transition
from the AFM$^1_{[110]}$ phase (where there is no stress by symmetry)
and the AFM$^2_{[110]}$ cubic phase with the same lattice constant. 
The transition produces a planar anisotropic stress, whose component 
$\tau_{[1\overline{1}0]}$ is tensile, i.e. the spin-antiparallel metal
atoms tend to get closer each other. In light of our results for FeN, we 
speculate that this magnetic order-derived stress is sensitive to
the magnitude of magnetization. Of course, the different bond length
(slightly larger for CrN than for FeN) plays a role as well and could 
justify in itself a different behavior.

\section{Electronic Properties}

\begin{figure}
\epsfxsize=8cm
\epsffile{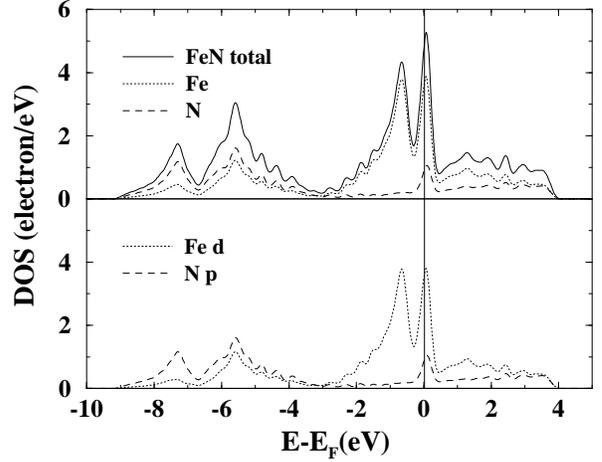}
\caption{Top panel: total and atom by atom decomposed DOS of PM NaCl-type
FeN. Bottom panel: Fe $d$-type and N $p$-type contributions to the total DOS.
In the shown energy window this two angular components are largely dominant.}
\label{dos_pm}
\end{figure}

In Figure \ref{dos_pm} the DOS for the PM phase of FeN is shown. In this
energy region, the total DOS (top panel) is almost completely
made up by d states of Fe and p states of N (bottom panel). The d-like
DOS shows two high peaks close to the Fermi energy (E$_F$). The lowest in 
energy is completely occupied and is composed by a mixture of 
75\% $t_{2g}$-type and 25\% $e_g$-type character. It is this peak which
falls at E$_F$ in paramagnetic CrN\cite{fpk} (a rigid band model works well
for the PM phase of this system). The second peak in the d DOS is entirely 
$t_{2g}$ in character, and is 
centered at E$_F$. The peaks located between -8 eV and -4 eV are mostly
N p, with a mixture of $t_{2g}$-type and (predominantly) $e_g$-type states.
The N p-like DOS gives an appreciable contribution to the DOS at the 
Fermi energy as well, unlike CrN for which only d-type states 
contribute. This DOS peak at E$_F$ arises from a strong $dp\sigma$
antibonding band. 

\begin{figure}
\epsfxsize=8cm
\epsffile{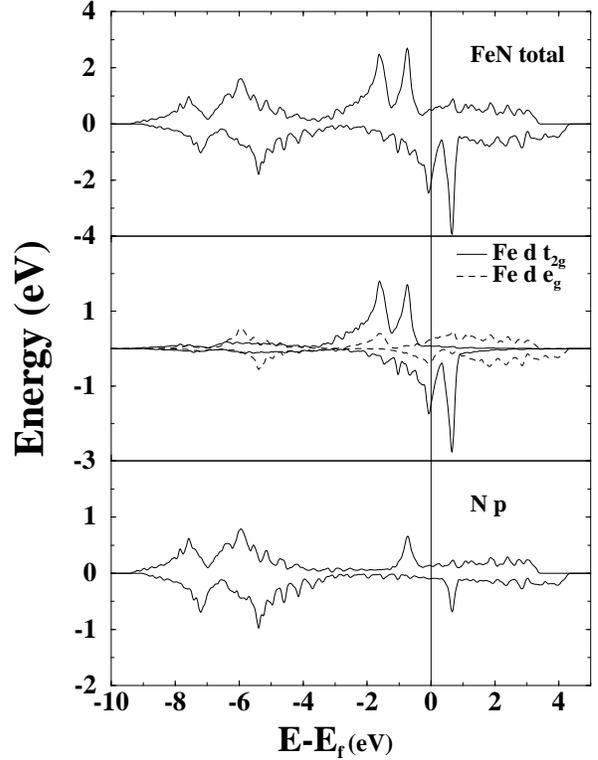}
\caption{Top panel: total DOS of FM NaCl-type FeN for spin up and down charge.
Middle: Fe $d$ $t_{2g}$-type (solid line) and  $e_g$-type (dashed)
contributions to DOS. Bottom: N $p$-type contribution.}
\label{dos_fm}
\end{figure}

The effect of FM ordering is easily visible in Figure \ref{dos_fm}.
Now the Fe d-like DOS (middle panel) is shown decomposed in $t_{2g}$ 
and  $e_g$ states.
The two d-type peaks are now split by an exchange splitting of roughly 1.6 eV.
After this splitting both peaks in the spin up channels are completely 
occupied, whereas the upper peak of the spin down channel is empty. 
Thus, 3 electrons per Fe atom saturate almost completely three 
$t_{2g}^{\uparrow}$ states, and $\sim$ 1.5 electron charge of mostly 
$t_{2g}$ type remains unpaired. Consistently with a cation to anion 
charge transfer of $\sim$ 1 electron, i.e. with a $d^7$ configuration for
Fe, there are 2.5 electrons occuping nearly paired $e_g$ states. 
The N p character in the DOS contributes to the magnetization, reflecting 
the N moment of $\sim$ 0.15 $\mu_B$. 

\begin{figure}
\epsfxsize=8cm
\epsffile{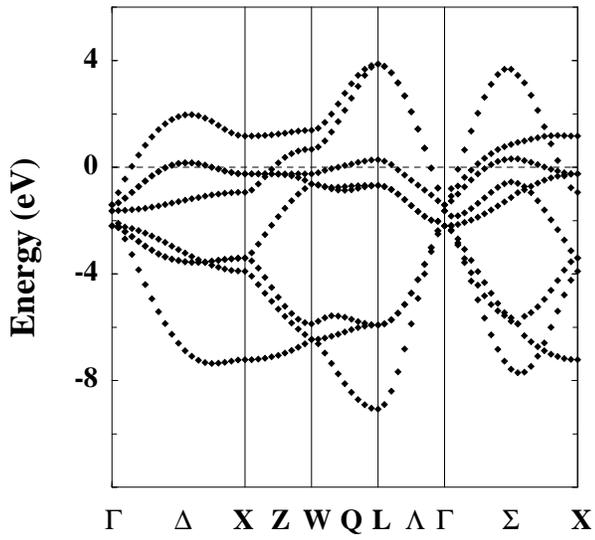}
\caption{Band structure for PM NaCl-type FeN. A flat band of $t_{2g}$
character running from $\Gamma$ to $L$ almost in touch with the Fermi
level is responsible for the pronounced peak of DOS at the Fermi Energy.}
\label{band_pm}
\end{figure}

\begin{figure}
\epsfxsize=8.5cm
\epsffile{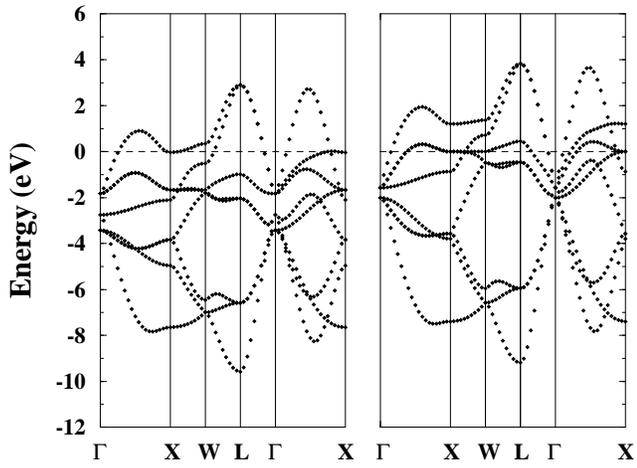}
\caption{Band structure for FM NaCl-type FeN. Left and right panels refer
to spin up and down bands, respectively. Directions and K-points are the
same of the previous Figure.}
\label{band_fm}
\end{figure}

The same features can be seen in the band structure.
In Figure \ref{band_pm} the PM FeN band structure is presented. 
It is easy to distinguish the very flat $t_{2g}$-type band 
crossing the Fermi level through $\Delta$, $Z$ and $Q$ directions, that
produces the half-occupied peak of PM DOS shown in Figure \ref{dos_pm}.
After FM ordering (Figure \ref{band_fm}), the band is split, in such
a way that the up band (left panel) is shifted $\sim$1.5-2 eV down to the 
Fermi energy, and the down band rises up slightly, but still remains
partially occupied. Evidently, much of the minority DOS peak at 
$\sim$0.8 eV arises from regions away from the symmetry lines plotted 
in Figure \ref{band_fm}. 

A feature observable in Figure \ref{band_fm} is that the spin splitting 
for some bands is strongly k-dependent, unlike the Stoner picture in which
the splitting is rigid. For example, all bands at $\Gamma$ are clustered
within 1 eV in the paramagnetic case and remain so in the minority bands, 
whereas in the majority bands they span a range of nearly 2 eV. Such 
non-rigid behavior has been observed in other transition metal compounds,
and reflects a combination of (i) an anisotropic exchange potential
(around the Fe atom) and (ii) hybridization between d orbitals on a 
strongly magnetic atom (Fe) and p orbitals on a weakly magnetic atom (N).

\section{Conclusions}

In this paper the structure of NaCl-type FeN has been investigated by 
ab initio calculations. The ferromagnetic phase is found to be stable 
against the paramagnetic and all the antiferromagnetic phases here considered. 
However, the energy differences between the magnetic phases are within
few hundredths of eV, and this fact, coupled with sensitivity of the magnetism
to stoichiometry, could explain the experimental variations reported
in the literature. In order to get insights
into the magnetic ordering mechanisms, a comparison with CrN was made.
Although for both compounds a sharp $d$-like peak is present at E$_F$,
the differences in charge transfer, magnetization and differing wavefunction
character at E$_F$ lead to very different magnetic properties. 
In particular, the orthorhombic 
distortion that is able to stabilize AFM CrN has no effect on
FeN. Since the distortion indicates the presence of tensile stress stored
on bonds between metal atoms with antiparallel spins, we suspect that the
stress could depend sensitively on the magnitude of magnetization.

The relief of stress is now accepted to be an important factor in the
structure and properties of surfaces and interfaces. Our results strongly 
suggest that the stress relief may be instrumental in determining the 
magnetic order in compounds such as FeN.

\section*{Acknowledgments}
This work was supported by National Science Foundation Grant No.
DMR-9802076. Computations were carried out at the San Diego
Supercomputing Center and at the Maui High Performance Computing
Center.

\newpage


\begin{table}
\caption{
Equilibrium lattice constant $a_0$ (in bohr), energy $E$ (in eV/formula unit)
and magnetic moment $m$ (in $\mu_B$/metal atom) of NaCl-type FeN
calculated for different magnetic phases. Energies are referred to the
lowest one, i.e. that of FM phase.
\label{struct}}
\begin{tabular}{cccccc}
        & FM   & PM    &  AFM$^1_{[110]}$ & AFM$^2_{[110]}$ & AFM$_{[111]}$ \\
\hline
$a_0$   & 7.55 & 7.52  & 7.55   & 7.52    &  7.52   \\
$E$     & 0    & 0.093 & 0.017  & 0.042   &  0.028  \\
\hline
$m$     & 1.65 &       & 1.44   & 1.46    &  1.15   \\
\end{tabular}
\end{table}


\end{document}